# Phase transition facilitated highly sensitive luminescence nanothermometry and thermal imaging

L. Marciniak[1*], W. Piotrowski[1], M. Szalkowski[1], V. Kinzhybalo[1],

M. Drozd[1], M. Dramicanin[2], A. Bednarkiewicz[1]

[1]Institute of Low Temperature and Structure Research, Polish Academy of Sciences, Okólna 2, 50-422 Wroclaw, Poland

[2] Vinca Institute of Nuclear Sciences – The National Institute of the Republic of Serbia, University of Belgrade, Serbia

* corresponding author: l.marciniak@intibs.pl

**Currently available temperature measurements or imaging at nano-micro scale are limited to fluorescent molecules and luminescent nanocrystals, whose spectral properties respond to temperature variation. The principle of operation of these conventional temperature probes is typically related to temperature induced multiphonon quenching or temperature dependent energy transfers, therefore, above 12%/K sensitivity and high thermal resolution remain a serious challenge. Here we demonstrate a novel class of highly sensitive thermographic phosphors operating in room temperature range with milikelvin thermal resolution, whose temperature readings are reproducible, luminescence is photostable and brightness is not compromised by thermal quenching. Corroborated with phase transition structural characterization and high spatio-temporal temperature imaging, we demonstrated that optically active europium ions are highly and smoothly susceptible to monoclinic to tetragonal phase transition in $LiYO_2$ host, which is evidenced by changed number and the splitting of Stark components as well as by smooth variation**

**of contribution between magnetic and electric dipole transitions. Further, reducing the size of phosphor from bulk to nanocrystalline matrix, shifted the phase transition temperature from 100ºC down to room temperature. These findings provide insights into the mechanism underlaying phase transition based luminescence nanothermometry and motivate future research toward new, highly sensitive, high temporal and spatial resolution nano-thermometers aiming at precise studying heat generation or diffusion in numerous biological and technology applications.**

Temperature and its precise quantification have been of interest to mankind for millennia [1,2]. This is mainly due to the fact that, this is one of the most important thermodynamic parameters, that affects and regulates rate and the character of the physical, chemical and biological processes or reactions [1,3–6]. Therefore, its accurate determination allows not only to passively understand the progress of the occurring processes and reactions but, most importantly, enables their reliable feedback-control [7–9]. For example, temperature is a valuable diagnostic parameter which enables recognition of the disease conditions like fever, inflammation spots, diabetic foot, cancerous tumors, heart strokes and many others [10–14]. Moreover, spatially and temporarily resolved temperature maps enable to study heat conduction properties in various materials in non-destructive and remote way[15]. The first ever discovered records regarding the concept of temperature evaluation was conceived by Philo of Byzantium from around the third century BC [1,2]. Since then, continuous work has been dedicated to propose new solutions to measure temperature in the most accurate, fastest and convenient way. In opposite to conventional thermometers, such as mercury based, thermocouples or bolometric detectors, a new and revolutionary approach, called luminescent thermometry, has been proposed [16–24]. It enables remote temperature readout by harnessing the thermally induced changes of the luminescent features of phosphor and thus opened new, previously unexplored possibilities for thermal measurements or imaging in highly corrosive environment [25], in high temperatures conditions[26], to sophisticated non-invasive *in vivo* and *in vitro* biological measurements through optical microscopes with unprecedented sub-micron size spatial and sub-second temporal resolutions[23,27–36]. In its most commonly used approaches, this technique exploits the susceptibility of emission spectra (e.g. ratio of the luminescence intensity of two emission bands) or the kinetics of luminescence lifetime of the excited level to the temperature[37–45]. The accuracy and reliability of the thermal imaging is, however, typically compromised by the probes of sufficient brightness and photostability, reproducibility of readout, simplicity of calibration, resistance to environmental factors and high relative sensitivity of the luminescent temperature probe to temperature changes[24,46]. As the susceptibility of luminescence intensity of the phosphor to temperature changes depends both on the host material and the electronic configuration of the optically active ions, the large-scale research has been focused on boosting the sensitivity of luminescent thermometers by optimization the type and concentration of active ions, modifying the phosphor host parameters such as phonon energy, crystal field strength and by involving various thermally induced interionic energy transfer processes[47–55]. Although the existing thermometers appear promising, their sensitivities values still leave room for further improvement[24,56,57]. Notably, an important limitation characterizing the vast majority of luminescence thermometers is the fact that the temperature range displaying highest sensitivity is typically correlating with temperature range where the intensity of luminescence is significantly quenched, thus limiting their brightness and thermometric performance[47,57,58]. Therefore, it is imperative to propose completely new solutions that enable temperature reading with high temperature sensitivities and resolutions, while maintaining high luminescence intensity.

In response to current limitations, and well defined requirements and expectations for luminescent thermometry field, here, we propose and evaluate a new method to determine temperature by exploiting the change of luminescence features in response to temperature-induced, first order crystal phase transitions. Importantly, these nanomaterials have been designed to quantify temperature in physiological range. The presented approach goes significantly beyond current paradigms in luminescence thermometry which exploit temperature-induced changes in the energy level populations and/or non-radiative transition rates. These latter methods are frequently compromised by the sensitivity of luminescence features (i.e. spectral fingerprints, intensity, quantum yield) to environmental factors other than temperature, inhomogeneous distribution of dopant ions, or contamination of luminescence probes[59]. The suitability of this newly proposed idea for highly sensitive luminescence thermometry is evidenced by a spectacular susceptibility of $Eu^{3+}$ ions, known to be luminescent structural probe, and its spectroscopic properties to a temperature induced continuous phase transition occurring to crystallographic environment and local site symmetries (Fig. 1). The $LiYO_2$, discussed here as a representative host material, reveals a phase transition from a monoclinic structure with *P*2$_1$/*c* symmetry to a tetragonal structure with *I*4/*amd* symmetry at a temperature of about 100 ºC in bulk materials[60–62]. However, by using materials of nanometric size, the phase transition temperature was purposefully reduced to room temperature (c.a. 25 ºC), which is of particular interest for biomedical microscopy studies. The phase transition induced by the temperature rise, not only changes the local symmetry of the ion, but reduces the number of Stark components to which the $^7F_1$ level is split, and increases their maximum splitting. These modifications lead to significant changes in the relative proportions of the $Eu^{3+}$ Stark components associated with the $^5D_0 \rightarrow {}^7F_1$ electronic transition, which are assigned to the two different crystallographic phases and corresponding different local $Eu^{3+}$ site symmetries. The spectacular changes of the luminescence intensity ratio (LIR) and the corresponding extraordinarily high values of the sensitivity have been observed in narrow, but biologically relevant temperature range. Simultaneously, the nano size phosphors satisfy the other strict requirements of biomedical applications, such as through optical microscope high spatial resolution detection and thermal imaging with relatively high temporal resolution.

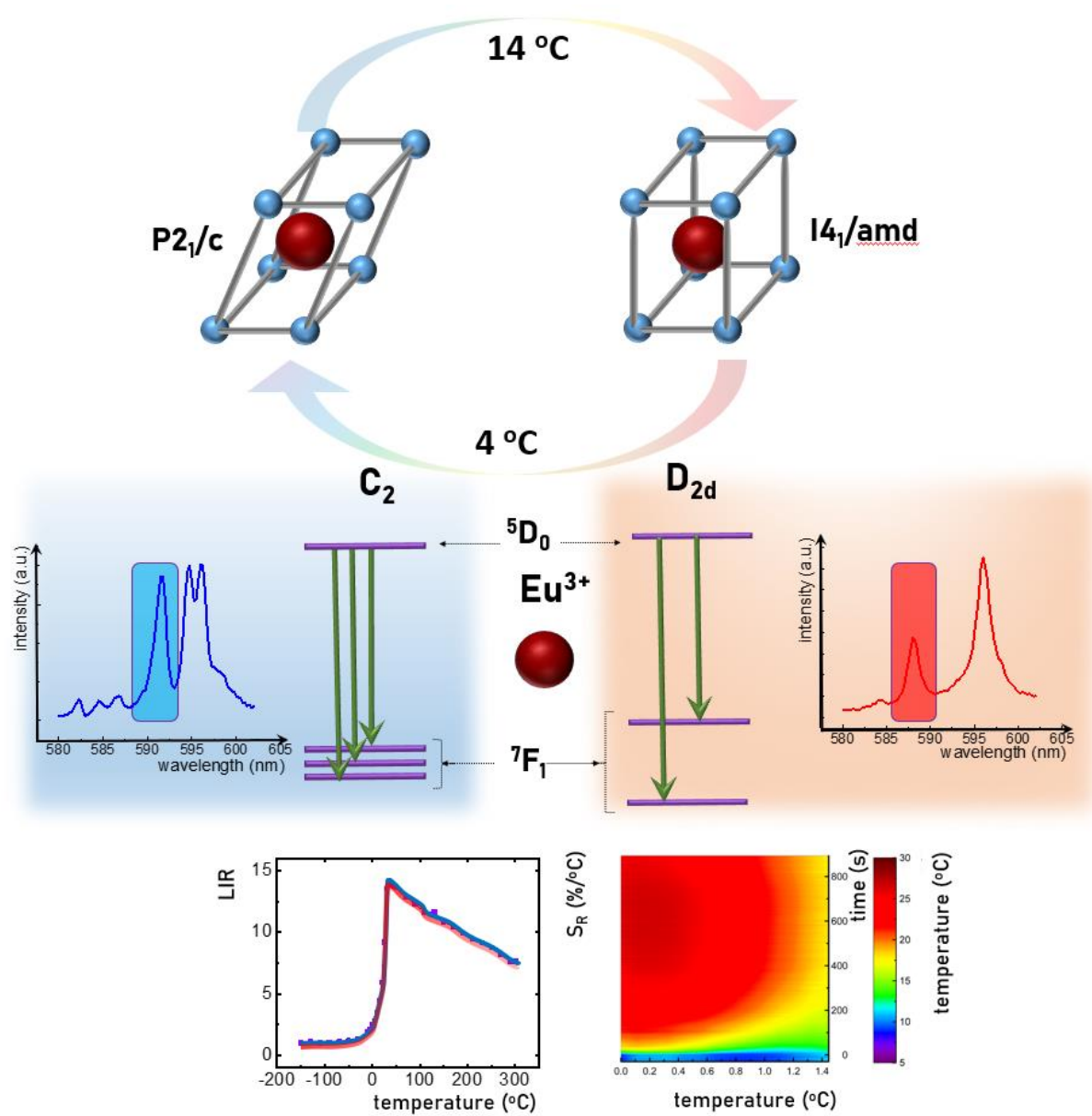

Figure 1. The conceptual image explaining the ideas of the proposed solution: the $LiYO_2$ phosphor undergoes the thermally induced, reversible, first-order phase transition from monoclinic structure of $P2_1/c$ space group to the tetragonal structure of $I4_1/amd$ space group at room temperature which affects the point symmetry of the luminescent $Eu^{3+}$ ion modifying its energy level diagram. Therefore the change of the temperature induces the quenching of the emission line at 590 nm and boosts the intensity of the 588 nm one. Their opposite thermal

dependence enables development of a highly sensitive luminescence-based remote temperature sensor of extraordinary relative sensitivity to temperature changes.

**RESULTS**

The group of host materials of a general formula Li$A$O$_2$ (where $A$ represents the rare earth ions) crystalizes in the three different structure types spanning from tetragonal of $I4_1/amd$ space group (Z=4), monoclinic $P2_1/c$ to orthorhombic of $Pbnm$ space group (Z=4), which is determined by the size of ionic radii of the $A$ ions. is LiYO$_2$. Blasse *et al.* reported high emission intensity of LiYO$_2$:Eu$^{3+}$ phosphor, which actually is one of the most promising and most efficient host material for luminescent application. Notably, in contrary to some other representatives of this family of compounds, LiYO$_2$ is characterized by low hygroscopic properties and is highly stable upon cathode ray excitation. The LiYO$_2$ crystallizes in the monoclinic crystal system of $a$ = 6.1493(8) Å, $b$ = 6.1500(10) Å, $c$ = 6.2494(2) Å and $β$ = 119.091(5)° cell parameters at low temperatures while undergoes a reversible phase transition to the tetragonal crystal system of $a$ = 4.4468(9) Å, $c$ = 10.372(22) Å cell parameters at higher temperatures (Figure 2a) [58]. In the case of the bulk material the first order phase transition was reported around 90 °C. However the lowering of the size of particles to the nanodimension enables reduction of the phase transition temperature to the room temperature range. The similarities in the ionic radii and the oxidation state facilitates the substitution of the octahedral Y$^{3+}$ site by the optically active lanthanide ions. The evidence of the structural phase transition can be confirmed by the measurement of the powder X-ray diffraction patterns as a function of temperature change (Figure 2b). Diffraction pattern taken at –50 °C indicates the presence of only LT phase. On heating the content of LT phase gradually decreases, whereas the HT phase appears, so both phases coexist in the temperature range of 0 - 25 °C. As shown in Fig. 2c, above 30 °C the reflections from LT phase disappear and the sample is pure HT form of LiYO$_2$ (the relative content of individual phases determined based on XRD patterns in the temperature function is presented Fig. S1). The differential scanning calorimetric curves clearly confirm that LiYO$_2$:Eu$^{3+}$ exhibits a reversible first-order phase transition (two-phase transitions with the noncontinuous character). The characteristic peaks or inflections for phase transitions are not observed in the -173 to 10 °C temperature range. During the heating a peak is observed with onset at 14.49 °C. The calculated enthalpies of this peak is equal to 9.43 J/g. During the cooling, the counterparts of observed peaks are noticed at 4.54 °C with calculated enthalpy – 7.3046 J/g, respectively. Additionally, the small temperature hysteresis is observed in the case of both peaks (ca. 10 °C for a maximum of peak between the heating and cooling cycle). The parameters of phase transition are stable and very similar in the subsequent cooling and heating cycles. The additional insight into the character of the phase transition provide temperature dependent Raman spectra (Fig.2 e and f). The factor group analysis indicates that only 9Ag+9Bg vibrations can be observed in Raman spectra for low temperature monoclinic phase On the other hand for the tetragonal LiYO$_2$ structure representation looks as follows: A1g+4A2u+4B1g+4Eu+4Eg+B2u, however only A1g+4B1g+4Eg are expected to be active in Raman spectrum. Therefore more rich spectra can be observed for the monoclinic structure and at elevated temperature above 20 °C the reduction of the number of vibrational peaks can be found (Figure 2e). This can be especially clearly seen is the 450-530 cm$^{-1}$ wavenumber range where the peaks intensity at 520 cm$^{-1}$ and 484 cm$^{-1}$ attributed to the bending of YO$_6$ group and stretching of O-Y-O gradually decrease at elevated temperatures. Moreover the position of the most intense peak at 859 cm$^{-1}$, is attributed to the stretching mode of YO$^{2-}$ group shifts toward higher wavenumbers (its shift as a function of temperature is presented in Fig. S2). The DLS measurements indicate that the sample consists of the particles of average grain size 54±15 nm (Fig.1g) what was also confirmed by the TEM images. The powder consists of highly crystalized and aggregated particles.

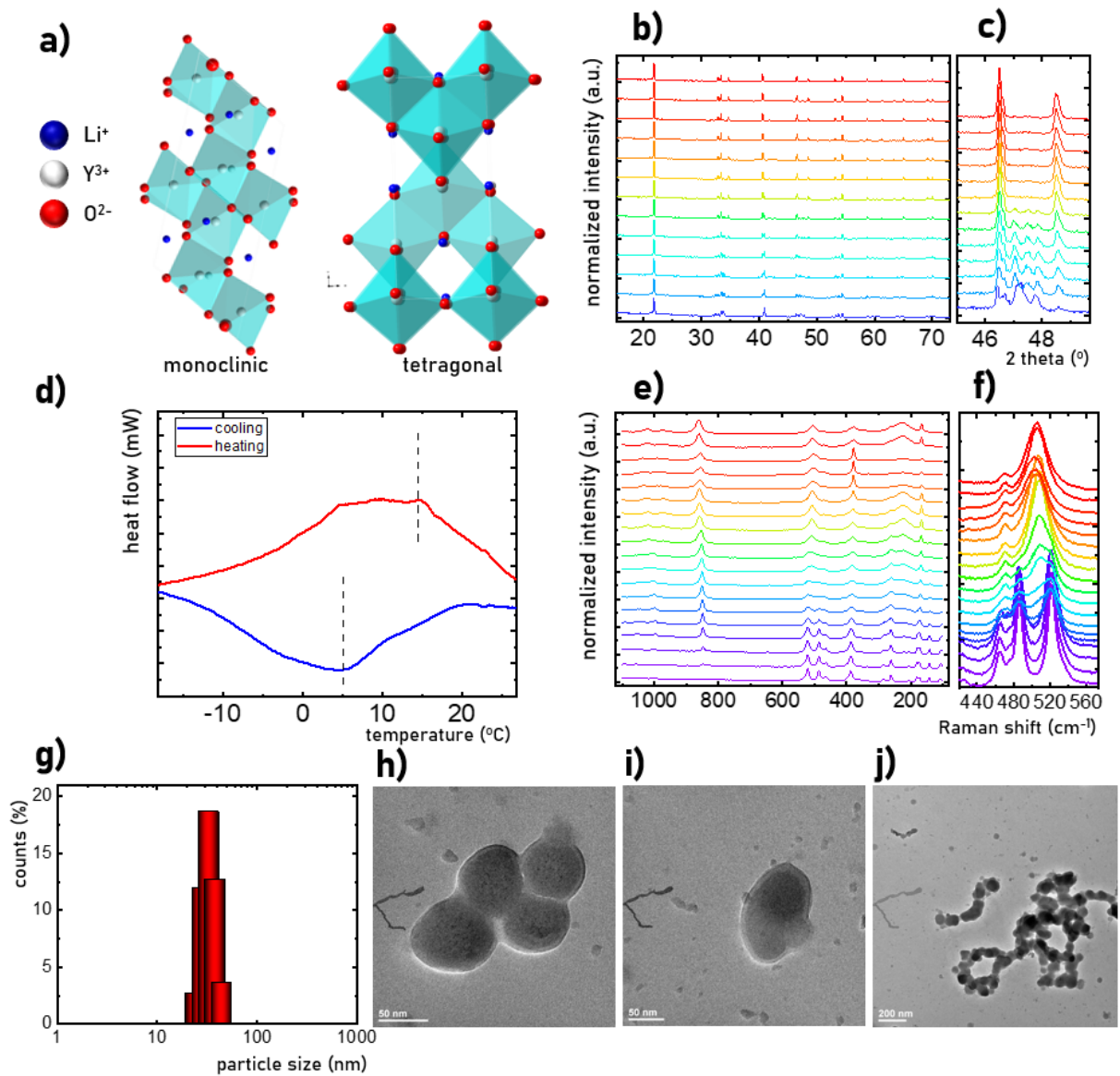

Figure 2. The visualization of the LiYO$_2$:Eu$^{3+}$ low temperature monoclinic and high temperature tetragonal crystal structures a), the thermal evolution of the XRD patterns in a 15-75° 2θ range -b) and zoom up in the 45-50° 2 θ range c). DSC curves obtained for LiYO$_2$:Eu$^{3+}$ phosphor d); thermal evolution of the Raman spectra of the LiYO$_2$:Eu$^{3+}$ e) and f); particle size distribution histograms obtained from DLS measurements g) and representative TEM images of the LiYO$_2$:Eu$^{3+}$ h-j).

The luminescent properties of the Eu$^{3+}$ ions doped phosphors are associated with the electronic depopulation of the metastable $^5D_0$ excited state to the lower lying $^7F_J$ multiplets, among which the electric dipole allowed $^5D_0 \rightarrow ^7F_2$ and magnetic dipole allowed $^5D_0 \rightarrow ^7F_1$ transitions are usually the most intense and are responsible for its red emission colour (Fig. S3). Therefore emission spectra of LiYO$_2$:Eu$^{3+}$ phosphor consist of two emission bands at around 590 and 620 nm associated with the $^5D_0 \rightarrow ^7F_1$ and $^5D_0 \rightarrow ^7F_2$ electronic transitions of Eu$^{3+}$ ions, respectively.

An increase of the temperature quench emission spectra, which can be attributed to the thermally activated energy migration to the nanoparticle surface. However, the most spectacular changes are associated with the thermally induced change of the shape of emission spectra, which is especially clearly visible in the case of the $^5D_0 \rightarrow {}^7F_1$ emission band (Figure 3b). At low temperature this band consists of three emission lines associated with the electronic transition to the three Stark sublevels of $^7F_1$ state expected to occur for the $C_2$ point symmetry. However in the case of the $D_{2d}$ point symmetry, which is expected for high temperature tetragonal phase of $LiYO_2$, the $^7F_1$ should be split into two components. Therefore, gradual increase of temperature results in the quenching of the lines at 586 nm (M(0→1)), 591 nm (M(0→2)) and 594 nm (M(0→3)) and the enhancement of the lines at 588 nm (T(0→1)) and 596 nm (T(0→2)). Moreover, slight c.a. 1 nm spectral shift of the M(0→2) and M(0→3) emission lines can be found in the investigated temperature range. The opposite thermal dependence of the lines attributed to the tetragonal and monoclinic crystal structures enables development of highly sensitive ratiometric luminescent thermometer (see Fig S4 for excitation spectra). However the spectral overlap of the M(0→3) and T(0→2) lines humpers the accurate remote temperature readout. Therefore the emission intensities of the T(0→1) and M(0→2) were analyzed (Figure 3c). At elevated temperatures the integral emission intensity M(0→2) gradually decreases up to -30 °C above which drastic quenching, up to 25% of the initial intensity, above 45 °C can be found. On the other hand above –30 °C, the intensity of T(0→1) increases 4-fold in the -30 °C to 35 °C temperature range, which is followed by the gradual reduction of its intensity associated with the thermal quenching of $Eu^{3+}$ luminescence. These unique and complex thermal trends enable to develop a favorable thermometric parameter (luminescence intensity ratio) as follows:

$$LIR = \frac{T(0 \rightarrow 1)}{M(0 \rightarrow 2)} = \frac{\int_{586.50nm}^{589.25nm} I(\lambda) dI}{\int_{589.66nm}^{589.75nm} I(\lambda) dI} \tag{1}$$

At elevated temperatures above -30 °C LIR becomes 5-fold enhanced at around 50 °C followed by a slight lowering of its value. The decrease of the LIR observed at higher temperatures is a reflection of the quenching of the T(0→1) emission line. It needs to be noticed that the change of the monotonicity of the thermal trend of LIR reduce the usable temperature range in which such luminescent thermal probe can be applied. The quantitative evaluation of the LIR enhancement and thus thermometric performance of the LiYO$_2$:Eu$^{3+}$ nanocrystals can be further quantified by calculating relative sensitivity of thermographic phosphors to temperature changes, which is expressed as:

$$S_R = \frac{1}{LIR}\frac{\Delta LIR}{\Delta T}100\% \qquad (2)$$

where ΔLIR represents a change of LIR value corresponding to the ΔT change of temperature. The sharp increase of the LIR below 50 °C is manifested as an unprecedently high value of the $S_R$ which reaches $S_R$=11.8%/°C at 22 °C and decreases when the temperature is further increased. The negative values of the $S_R$ above 50 °C corresponds to the change of the thermal monotonicity of the LIR. The hysteresis of the temperatures at which phase transitions occur implies the necessity of the verification of the repeatability of the temperature readout using LiYO$_2$:Eu$^{3+}$ as a thermal probe. The presented evaluation of the LIR values within several heating-cooling cycles between -50 and 50 °C reveals the remarkably good thermometric performance of the considered phosphor. The detailed analysis of the thermal evolution of LIR within heating-cooling cycles, however, indicated that the hysteresis of the LIR value can be found (Figure S5 and S6). The measurement of the hysteresis loop within heating cooling cycles indicates an excellent repeatability of the LIR values within the change of the temperature with the same monotonicity. This indicates that the LiYO$_2$:Eu$^{3+}$ phosphor can be used for temperature determination in the application where monotonic change of the temperature can be found.

The accuracy of the remote thermal imaging depends on the uncertainty of temperature determination which can be calculated as follows

$$\delta T = \frac{1}{S_R} \frac{\delta LIR}{LIR} \qquad (3)$$

where the δLIR is the standard deviation determined based on the 30 measurements of the emission spectra at constant temperature. In the case of the LiYO$_2$:Eu$^{3+}$, δT decreases with temperature reaching δT=0.005 °C at 49 °C (Figure S7)

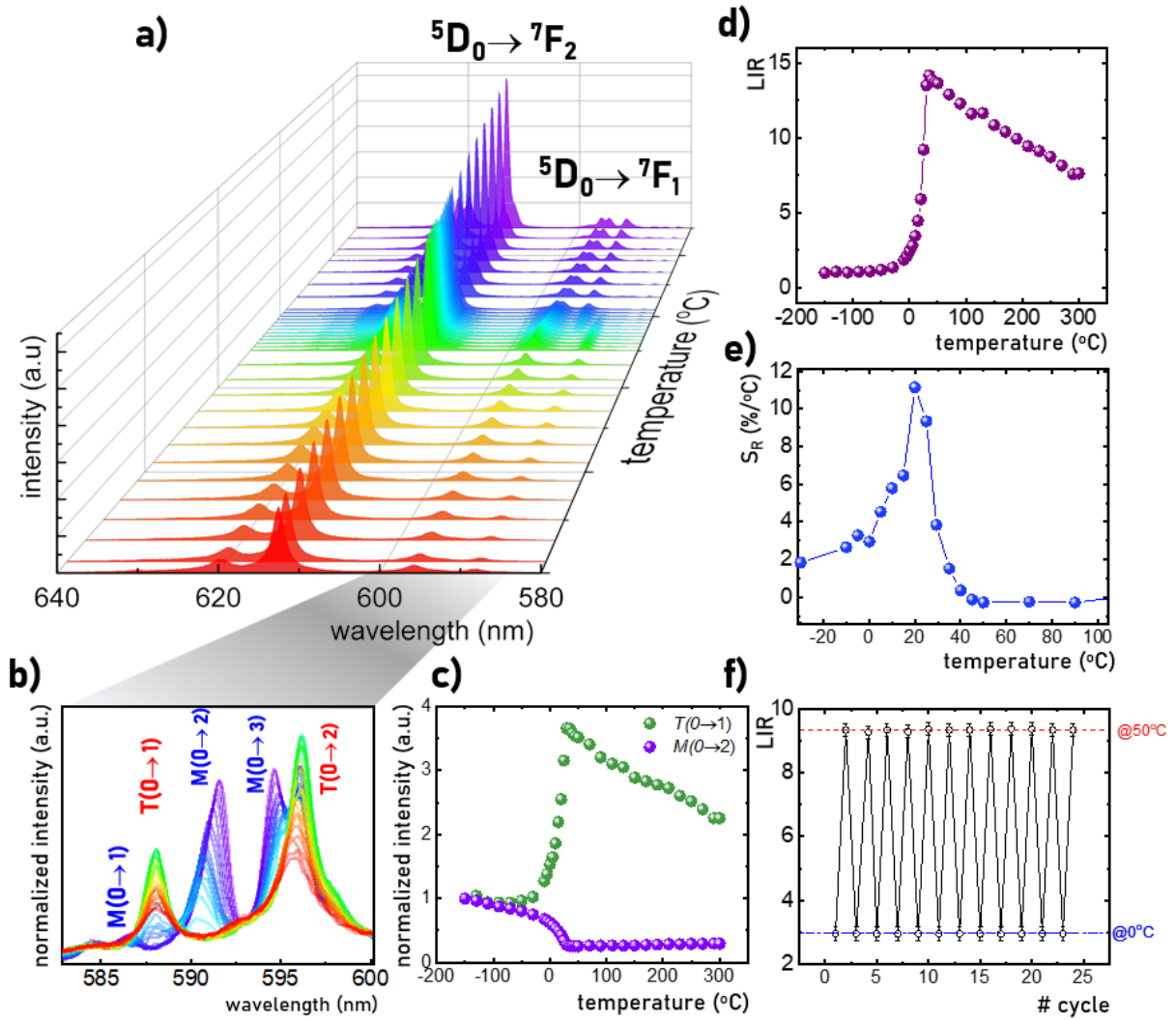

Figure 3. Thermal evolution of the emission spectra of LiYO$_2$:Eu$^{3+}$ phosphor (a); the influence of the temperature on the shape of the $^5D_0 \rightarrow {}^7F_1$ emission band (b); integral emission intensities of T(0→1) and

M(0→2) lines (c) and their luminescence intensity ratio (d) measured as a function of temperature; thermal dependence of $S_R$ (e) and the LIR measured within several heating-cooling cycles (f).

Although thermocouple or termovision cameras enable temperature monitoring with similar temporal and temperature resolution to the one presented here, none of existing techniques is capable to offer sufficiently high spatial resolution (here c.a. 0.1 mm) and temperature visualization under optical microscope. Therefore, luminescent nanoparticle based temperature sensing, which enable combing sub-second temporal resolution with sub-millimeter spatial resolution, give promises for many interesting applications in (bio)technology. To demonstrate the suitability of such nanothermometers for optical microscopy, we have performed spatially and temporarily resolved studies of a point heat source.

The sample composed of $LiYO_2$:1% $Eu^{3+}$ deposited on bare glass substrate was attached (with good thermal contact) to the cooper cooling plate equipped with the set of two thermoelectric Peltier modules and heat dissipating system to provide the sample initial temperature of 30 °C (Fig. S8). On the other hand, another Peltier module based system with cooper finger responsible for local heating of the sample. Such setup, composed of two modules, heating and cooling ones, was mounted on the microscopic stage and adjusted in a configuration enabling observation on the microscope and recording the image of the luminescence corresponding to some selected area of the sample (along L=1.5mm path), with the heating finger edge positioned on one side of the imaged area (Fig. S9-11).

As soon as thermal stabilization of the sample was achieved with cooling plate, collection of the consecutive frames was initialized, where each of them was containing information about the local luminescence spectrum of the sample along the observed area. After initial 30 s of the observation of the sample kept in the steady state conditions, the heating system was turned on and local heating of the sample has begun.

The set of spectra measured in consecutive time frames in the central part of the sample is presented in the Fig. 4.(a). This temporal evolution of the luminescence spectrum exhibits two opposite effects for the two spectral bands being analyzed. For the M(0→2) band (marked in blue), after the first 30 s of the observation characterized by absence of any substantial changes of the intensity, a gradual quenching starting at the $t_0 = 0$ s can be observed. Oppositely, in the case of the T(0→1) band (marked in red) one can observe gradual increasing of the intensity over time. Because these effects are opposite to each other, their relation can be considered as a sensitive temperature indicator. We have therefore proposed a way to translate this spectral information, obtained for given spatial position and time frame into the information about corresponding temperature (see SI). By performing calibration procedure, based on relation between intensities determined for the T(0→1) and M(0→2) bands, being corrected by the background signal measured away from the analyzed peaks (Bkgd, marked in green) and followed by automatic software analysis of the 4D data structure, temporal evolution of the temperature along the sample was calculated (Fig. 4(d)).

The thermal profiles were fitted to the model accounting for the transient thermal dissipation via convection to the environment perpendicular to the plate and the steady-state heat diffusion through the plate[63]:

$$T(x,t) = T_a - \frac{\Delta T}{\cosh(m \cdot x)} \cdot \exp(-t/\tau), \qquad (4)$$

where $\Delta T$ is the temperature rise due to the constant flux thermal source, $T_a$ is the ambient temperature, $\tau = R_t C_t$ ($R_t$ is the convection resistance and $C_t$ is the lumped thermal capacitance of the plate), and $m = (4h/kD)^{1/2}$ where $h$ is the heat transfer coefficient, $k$ is the thermal conductivity of the plate, and $D$ is the width of the plate. Perfect correlation between experimental data and the theoretical model confirms the high accuracy of remote temperature determination using LiYO$_2$:Eu$^{3+}$ phosphor. Because our point heat source and the sample is cylindrically symmetric, we have studied heat diffusion process along one direction, with the benefit of simultaneous kinetics visualization, which ultimately enables to judge on heat conduction of the substrate materials. This fit-for-purpose demonstration can be easily extended to two dimensional spatially and temporarily resolved imaging.

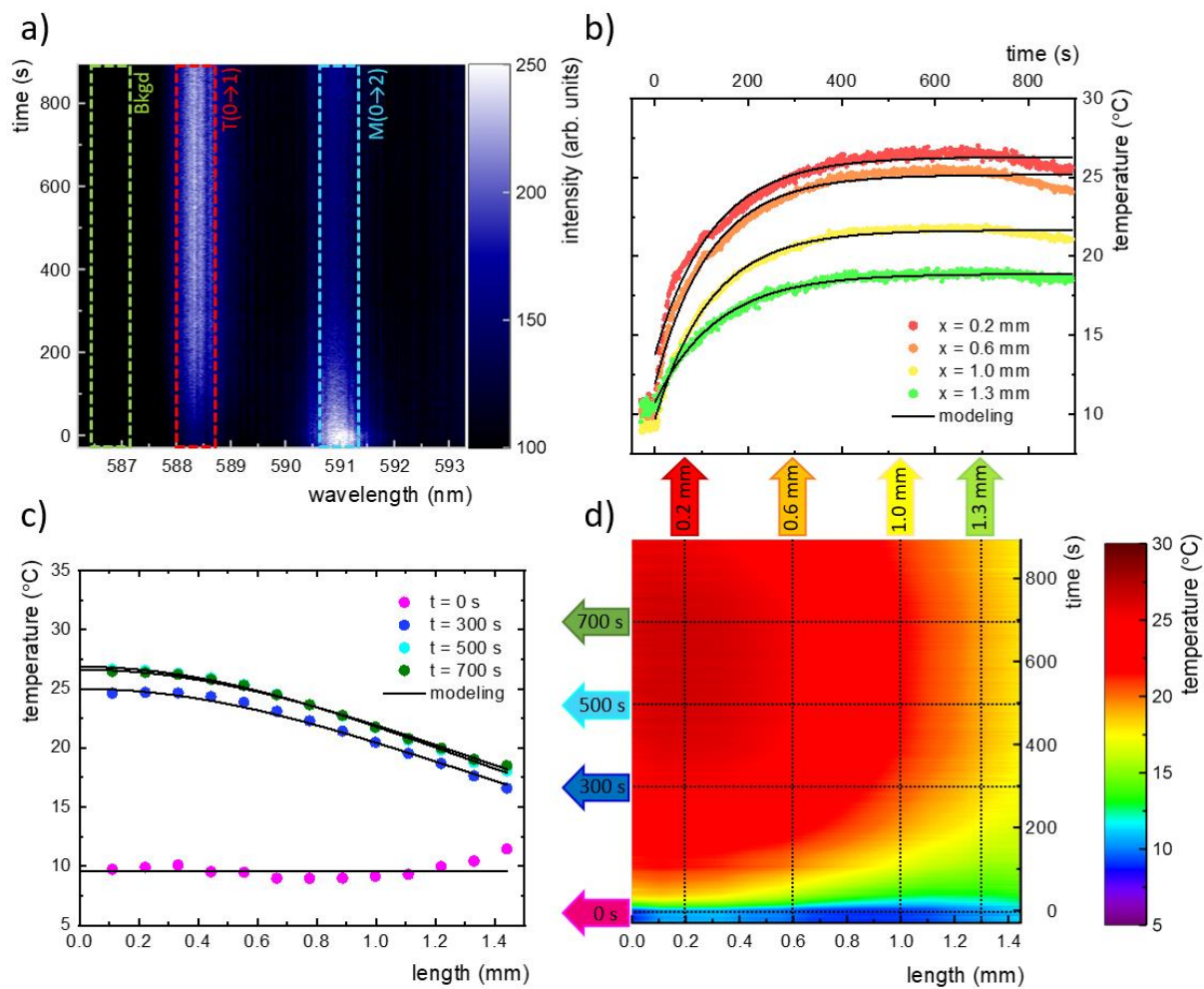

**Figure 4.** Microscopic measurements of the spatial and temporal temperature distribution around point heat source. (a) Emission spectra measured at the consecutive moments after turning on localized heating; the spectral ranges of the T(0→1) M(0→2), and the background signal level are marked with red, blue and green regions, respectively. (d) Temporal (y axis) evolution of the spatial temperature distribution (x axis) along the sample from the 'hot finger' away towards cooling plate. The black vertical lines represent the positions of the profiles are showing temporal changes of the temperature at selected positions along the sample (b), while the black horizontal

lines represent the profiles showing spatial temperature distribution at the selected moments since the start of the heating (c).

In fact, such temperature distribution in space and in time provides comprehensive information about dynamics of the temperature gradients induced by the heating finger present at the left side of the sample (position length=0 mm). Thus, such dynamics can by analyzed from the two perspectives – (1) as the temporal evolution of the temperature at various positions along the sample, or, alternatively, (2) as the temperature distribution at the selected moments in time. Both of these analyses can be performed utilizing the profiles taken in proper dimension from the whole dataset. Selected profiles corresponding to both of these approaches are shown by the black dotted lines in the Fig. 4(d). The set of profiles marked by the vertical lines and presenting the temporal temperature evolution at the given positions (0.2 mm, 0.6 mm, 1.0 mm and 1.3 mm) is presented in Fig. 4(b). The common structure can be found for all of such extracted profiles, namely they start from similar and stable initial temperature, however as soon as the heating is turned on, the temperature starts to grow. Moreover, the growth dynamics vary between the positions – the closer to the heater the more dynamic the temperature increase is.

Analogously, the complementary dataset composed of the profiles extracted along the horizontal lines (Fig. 4(c)) present the temperature distribution along the sample at various moments in time (0 s, 100 s, 300 s, 600 s). Initially the temperature along the sample is around 10 °C, however, once the heating starts, clear trend can be observed – the temperature is growing with time, nevertheless this process is not homogeneous in space. The temperature increase is more dynamic in close vicinity to the heater position (~0.0 – 0.3 mm) - here the temperature is also the highest. At longer distances away from the hot finger, the process is getting less dynamic and observed temperatures are lower. Furthermore, the dynamics changes also with time – after first, quite rapid changes, the system is going toward the saturation and the observed increase of the temperature is slowing down.

**CONCLUSIONS**

The constant need to increase the accuracy of remote temperature reading and mapping justifies the continuous interest in new materials with unique thermometric properties that can meet these requirements. In this work a strategy that involves first order phase transition between monoclinic structure of *P*2$_1$/*c* space group (low temperature phase) to the tetragonal structure of *I*4$_1$/*amd* space group (high temperature phase) in LiYO$_2$:Eu$^{3+}$ nanocrystals is proposed. Thermally induced change of the structural phase and thus local Eu$^{3+}$ point symmetry initiate significant changes in its luminescent properties, that are especially clearly manifested in the change of the $^5D_0 \rightarrow ^7F_1$ emission band. Therefore, the intensity ratio of emission lines associated with the transition from $^5D_0$ state to two different Stark components of $^7F_1$ level was used as a thermometric parameter. The value of this parameter increases at temperature elevation up to 50 °C, what indicates its excellent thermometric performance. The unprecedentedly high relative sensitivity $S_R$=11.8%/°C and 22 °C with the favorably low temperature determination uncertainty $\delta T$=0.005 °C, confirm that LiYO$_2$:Eu$^{3+}$ is one of the most promising materials for luminescent-based temperature determination. The observed hysteresis loop of thermal dependence of LIR values, with high repeatability of its value within the change of temperature of one, defined monotonicity suggest that LiYO$_2$ can be used as a temperature probe in the applications where one monotonicity of temperature change is expected to occur. Therefore, in the proof of the concept experiment, the LiYO$_2$:Eu$^{3+}$ was used to study heat distribution with sub-millimeter spatial resolution and millisecond temporal resolution in which thermal hysteresis of LIR dependence will not provide any limitations. The presented results suggest that the luminescent materials displaying first order phase transition may, in general, constitute a new promising direction of the research targeted to the development of highly sensitive thermographic phosphors and open new possibilities of thermal imaging with millikelvin thermal resolution.

## ACKNOWLEDGEMENTS

M.Sz. and A.B. acknowledge financial support from 2018/31/B/ST5/01827.

## REFERENCES:

1. McGee, T. D. *Principles and Methods of Temperature Measurement*. (Wiley-Interscience; 1st edition, 1988).

2. Dramićanin, M. Chapter 1 - Introduction to Measurements of Temperature. in *Woodhead Publishing Series in Electronic and Optical Materials* (ed. Dramićanin, M. B. T.-L. T.) 1–12 (Woodhead Publishing, 2018). doi:https://doi.org/10.1016/B978-0-08-102029-6.00001-4

3. Chang Hasok. *Inventing Temperature: Measurement and Scientific Progress*. (Oxford University Press, 2004).

4. Sonna, L. A., Fujita, J., Gaffin, S. L. & Lilly, C. M. Invited Review: Effects of heat and cold stress on mammalian gene expression. *Journal of Applied Physiology* **92**, 1725–1742 (2002).

5. McMahon, H. T. & Boucrot, E. Molecular mechanism and physiological functions of clathrin-mediated endocytosis. *Nature Reviews Molecular Cell Biology* **12**, 517 (2011).

6. Baffou, G., Rigneault, H., Marguet, D. & Jullien, L. A critique of methods for temperature imaging in single cells. *Nature Methods* **11**, 899–901 (2014).

7. Kok, H. P. *et al.* Heating technology for malignant tumors: a review. *International Journal of Hyperthermia* **37**, 711–741 (2020).

8. Jha, S., Sharma, P. K. & Malviya, R. Hyperthermia: Role and Risk Factor for Cancer Treatment. *Achievements in the Life Sciences* **10**, 161–167 (2016).

9. Otte, J. Hyperthermia in cancer therapy. *European Journal of Pediatrics* **147**, 560–569 (1988).

10. van Netten, J. J. *et al.* Diagnostic Values for Skin Temperature Assessment to Detect Diabetes-Related Foot Complications. *Diabetes Technology & Therapeutics* **16**, 714–721 (2014).

11. Casas-Alvarado, A. *et al.* Advances in infrared thermography: Surgical aspects, vascular changes, and pain monitoring in veterinary medicine. *Journal of Thermal Biology* **92**, 102664 (2020).

12. Deng, Z.-S. & Liu, J. Mathematical modeling of temperature mapping over skin surface and its implementation in thermal disease diagnostics. *Computers in Biology and Medicine* **34**, 495–521 (2004).

13. S. Maya, Bruno Sarmento, Amrita Nair, N. Sanoj Rejinold, Shantikumar V. Nair, R. J. Smart Stimuli Sensitive Nanogels in Cancer Drug Delivery and Imaging: A Review. *Current Pharmaceutical Design* **19**, (2013).

14. Rivens, I., Shaw, A., Civale, J. & Morris, H. Treatment monitoring and thermometry for therapeutic focused ultrasound. *International Journal of Hyperthermia* **23**, 121–139 (2007).

15. Reparaz, J. S. *et al.* A novel contactless technique for thermal field mapping and thermal conductivity determination: Two-Laser Raman Thermometry. *Review of Scientific Instruments* **85**, 34901 (2014).

16. Bradley, L. C. A Temperature-Sensitive Phosphor Used to Measure Surface Temperatures in Aerodynamics. *Review of Scientific Instruments* **24**, 219–220 (1953).

17. Zhou, J., del Rosal, B., Jaque, D., Uchiyama, S. & Jin, D. Advances and challenges for fluorescence nanothermometry. *Nature Methods* **17**, 967–980 (2020).

18. Dramićanin, M. Chapter 5 - Methods of Analysis for Luminescence Thermometry Measurements. in *Woodhead Publishing Series in Electronic and Optical Materials* (ed. Dramićanin, M. B. T.-L. T.) 85–112 (Woodhead Publishing, 2018). doi:https://doi.org/10.1016/B978-0-08-102029-6.00005-1

19. Jaque, D. & Vetrone, F. Luminescence nanothermometry. *Nanoscale* **4**, 4301–4326 (2012).

20. Wang, X., Wolfbeis, O. S. & Meier, R. J. Luminescent probes and sensors for temperature. *Chemical Society Reviews* **42**, 7834–7869 (2013).

21. Brites, C. D. S. *et al.* Thermometry at the nanoscale. *Nanoscale* **4**, 4799 (2012).

22. Brites, C. D. S. S., Millán, A. & Carlos, L. D. Chapter 281 - Lanthanides in Luminescent Thermometry. in *Including Actinides* (eds. Jean-Claude, B. & Vitalij K., P. B. T.-H. on the P. and C. of R. E.) **49**, 339–427 (Elsevier, 2016).

23. Vetrone, F. *et al.* Temperature sensing using fluorescent nanothermometers. *ACS Nano* **4**, 3254–3258 (2010).

24. Bednarkiewicz, A., Marciniak, L., Carlos, L. D. & Jaque, D. Standardizing luminescence nanothermometry for biomedical applications. *Nanoscale* **12**, 14405–14421 (2020).

25. Allison, S. W. & Gillies, G. T. Remote thermometry with thermographic phosphors: Instrumentation and applications. *Review of Scientific Instruments* **68**, 2615–2650 (1997).

26. Khalid, H. A. & Kontis, K. Thermographic Phosphors for High Temperature Measurements: Principles, Current State of the Art and Recent Applications. *Sensors* **8**, 5673–5744 (2008).

27. Chrétien, D. *et al.* Mitochondria are physiologically maintained at close to 50 °C. *PLOS Biology* **16**, e2003992 (2018).

28. Hong, G. *et al.* Through-skull fluorescence imaging of the brain in a new near-infrared window. *Nature Photonics* **8**, 723–730 (2014).

29. Yang, Y. *et al.* Fluorescent N-Doped Carbon Dots as in Vitro and in Vivo Nanothermometer. *ACS Applied Materials and Interfaces* **7**, 27324–27330 (2015).

30. Arai, S. *et al.* Mitochondria-targeted fluorescent thermometer monitors intracellular temperature gradient. *Chemical Communications* **51**, 8044–8047 (2015).

31. Kiyonaka, S. *et al.* Validating subcellular thermal changes revealed by fluorescent

thermosensors. *Nature Methods* **12**, 801–802 (2015).

32. Kucsko, G. *et al.* Nanometre-scale thermometry in a living cell. *Nature* **500**, 54 (2013).

33. Carrasco, E. *et al.* Intratumoral Thermal Reading During Photo-Thermal Therapy by Multifunctional Fluorescent Nanoparticles. *Advanced Functional Materials* **25**, 615–626 (2015).

34. Gota, C., Okabe, K., Funatsu, T., Harada, Y. & Uchiyama, S. Hydrophilic Fluorescent Nanogel Thermometer for Intracellular Thermometry. *Journal of the American Chemical Society* **131**, 2766–2767 (2009).

35. Cheng, L. *et al.* PEGylated WS2 Nanosheets as a Multifunctional Theranostic Agent for in vivo Dual-Modal CT/Photoacoustic Imaging Guided Photothermal Therapy. *Advanced Materials* **26**, 1886–1893 (2014).

36. Hemmer, E., Acosta-Mora, P., Méndez-Ramos, J. & Fischer, S. Optical nanoprobes for biomedical applications: shining a light on upconverting and near-infrared emitting nanoparticles for imaging, thermal sensing, and photodynamic therapy. *J. Mater. Chem. B* **5**, 4365–4392 (2017).

37. del Rosal, B. *et al.* Infrared-Emitting QDs for Thermal Therapy with Real-Time Subcutaneous Temperature Feedback. *Advanced Functional Materials* **26**, 6060–6068 (2016).

38. Qiu, X. *et al.* Ratiometric upconversion nanothermometry with dual emission at the same wavelength decoded via a time-resolved technique. *Nature Communications* **11**, 4 (2020).

39. Xu, M. *et al.* Ratiometric nanothermometer in vivo based on triplet sensitized upconversion. *Nature Communications* **9**, (2018).

40. Pickel, A. D. *et al.* Apparent self-heating of individual upconverting nanoparticle thermometers. *Nature Communications* **9**, 4907 (2018).

41. Liu, H. *et al.* Intracellular Temperature Sensing: An Ultra-bright Luminescent Nanothermometer with Non-sensitivity to pH and Ionic Strength. *Scientific Reports* **5**, 14879

(2015).

42. Yakunin, S. *et al.* High-resolution remote thermometry and thermography using luminescent low-dimensional tin-halide perovskites. *Nature Materials* **18**, 846–852 (2019).

43. Zhu, X. *et al.* Temperature-feedback upconversion nanocomposite for accurate photothermal therapy at facile temperature. *Nature Communications* **7**, 10437 (2016).

44. Bao, G., Wong, K.-L., Jin, D. & Tanner, P. A. A stoichiometric terbium-europium dyad molecular thermometer: energy transfer properties. *Light: Science & Applications* **7**, 96 (2018).

45. Ye, F. *et al.* Ratiometric Temperature Sensing with Semiconducting Polymer Dots. *Journal of the American Chemical Society* **133**, 8146–8149 (2011).

46. Bednarkiewicz, A. *et al.* Luminescence based temperature bio-imaging: Status, challenges, and perspectives. *Applied Physics Reviews* **8**, 11317 (2021).

47. Suta, M. & Meijerink, A. A Theoretical Framework for Ratiometric Single Ion Luminescent Thermometers—Thermodynamic and Kinetic Guidelines for Optimized Performance. *Advanced Theory and Simulations* **3**, 2000176 (2020).

48. Suta, M. *et al.* Making Nd3+ a Sensitive Luminescent Thermometer for Physiological Temperatures—An Account of Pitfalls in Boltzmann Thermometry. *Nanomaterials* **10**, 543 (2020).

49. Kniec, K., Ledwa, K. A., MacIejewska, K. & Marciniak, L. Intentional modification of the optical spectral response and relative sensitivity of luminescent thermometers based on Fe3+,Cr3+,Nd3+co-doped garnet nanocrystals by crystal field strength optimization. *Materials Chemistry Frontiers* **4**, 1697–1705 (2020).

50. Brik, M. G., Papan, J., Jovanović, D. J. & Dramićanin, M. D. Luminescence of Cr3+ ions in ZnAl2O4 and MgAl2O4 spinels: Correlation between experimental spectroscopic studies and crystal field calculations. *Journal of Luminescence* **177**, 145–151 (2016).

51. del Rosal, B. *et al.* Neodymium-Based Stoichiometric Ultrasmall Nanoparticles for

Multifunctional Deep-Tissue Photothermal Therapy. *Advanced Optical Materials* **4**, 782–789 (2016).

52. Matuszewska, C. & Marciniak, L. The influence of host material on NIR II and NIR III emitting Ni2+-based luminescent thermometers in ATiO3: Ni2+ (A = Sr, Ca, Mg, Ba) nanocrystals. *Journal of Luminescence* **223**, 117221 (2020).

53. Marciniak, L. & Trejgis, K. Luminescence lifetime thermometry with $Mn^{3+}$-$Mn^{4+}$co-doped nanocrystals. *Journal of Materials Chemistry C* **6**, (2018).

54. Elzbieciak, K., Bednarkiewicz, A. & Marciniak, L. Temperature sensitivity modulation through crystal field engineering in Ga3+ co-doped Gd3Al5-xGaxO12:Cr3+, Nd3+ nanothermometers. *Sensors and Actuators B: Chemical* **269**, 96–102 (2018).

55. del Rosal, B. *et al.* Neodymium-doped nanoparticles for infrared fluorescence bioimaging: The role of the host. *Journal of Applied Physics* **118**, 143104 (2015).

56. Uchiyama, S. & Gota, C. Luminescent molecular thermometers for the ratiometric sensing of intracellular temperature. *Reviews in Analytical Chemistry* **36**, 20160021 (2017).

57. Dramićanin, M. D. Trends in luminescence thermometry. *Journal of Applied Physics* **128**, 40902 (2020).

58. Bednarkiewicz, A., Marciniak, L., Carlos, L. D. & Jaque, D. Standardizing luminescence nanothermometry for biomedical applications. *Nanoscale* **12**, 14405–14421 (2020).

59. Reisfeld* †, R., Zigansky, E. & Gaft, M. Europium probe for estimation of site symmetry in glass films, glasses and crystals. *Molecular Physics* **102**, 1319–1330 (2004).

60. Faucher, M. D., Sciau, P., Kiat, J.-M., Alves, M.-G. & Bouree, F. Refinement of the Monoclinic and Tetragonal Structures of Eu3+-Doped LiYO2by Neutron Diffraction at 77 and 383 K Differential Scanning Calorimetry, and Crystal Field Analysis. *Journal of Solid State Chemistry* **137**, 242–248 (1998).

61. Faucher, M. D. *et al.* Optical and Crystallographic Study of Eu3+and Tb3+Doped LiYO2:

Phase Transition, Luminescence Efficiency and Crystal Field Calculation. *Journal of Solid State Chemistry* **121**, 457–466 (1996).

62. Kyoung Moune, O., Dexpert-Ghys, J., Piriou, B., Marie-Gabrielle Alves & Faucher, M. D. Electronic structure of Pr3+ and Tm3+ doped LiYO2. *Journal of Alloys and Compounds* **275–277**, 258–263 (1998).

63. Sadık Kakac, Yaman Yener, C. P. N.-C. *Heat Conduction*. (CRC Press, 2018).

# METHODS

*Synthesis*

The powders of LiYO$_2$:1% Eu$^{3+}$ nanocrystals were synthesised with a modified Pechini method. Li$_2$CO$_3$ (99.9% purity, Chempur), Y$_2$O$_3$ (99.999% purity, Stanford Materials Corporation), Eu$_2$O$_3$ (99.99% purity, Stanford Materials Corporation), C$_6$H$_8$O$_7$ (>99.5% purity, Alfa Aesar) and H(OCH$_2$CH$_2$)$_n$OH, (PEG-200, n = 200, Alfa Aesar) were used as starting materials. Yttrium and europium oxides were dissolved in deionized water with the addition of a small amount of HNO$_3$ (65% purity, Avantor), then recrystallized three times to remove the excess of nitrogen. The 2-, 3- or 4-fold stoichiometric excess of lithium carbonate were added to the water solution of nitrates. After that, an anhydrous citric acid and polyglycol were added to the mixture. The molar ratio of citric acid to all metals was set up as 6:1, meanwhile PEG-200 and citric acid were used in a molar ratio of 1:1. Subsequently, the obtained solution was dried for 24 hours at 125 °C until a resin was formed. The produced resin of the samples with 1% Eu$^{3+}$ concentration in respect to the number of Y$^{3+}$ moles ions was annealed in porcelain crucibles for 6 h in air at a temperature of 800, 850, 900 and 1000 °C.

*Characterization*

Powder diffraction data were obtained using a PANalytical X'Pert Pro diffractometer equipped with Oxford Cryosystems Cryostream 700 Series cooler using Ni-filtered Cu Kα radiation ($V = 40$ kV, $I = 30$ mA). The sample was put into Hampton Research glass capillary and sealed. Diffraction patterns in 15-90° 2θ range were measured in heating/cooling sequence in the temperature range of –50 to +75 °C (from 0 to 50 °C every 5 °C). ICSD database entries No. 50992 (LT phase) and 50993 (HT phase) were taken as initial models for the analysis of the obtained diffraction data. Transmission electron microscope (TEM) images were performed with the Philips CM-20 SuperTwin transmission electron microscope, operating at 160 kV. The sample was ground in a mortar and dispersed in methanol, and then a drop of the suspension was put on a copper microscope grid covered with carbon. Before the measurement, the sample was dried and purified in a $H_2/O_2$ plasma cleaner for 1 min. A differential scanning calorimetric (DSC) measurements were performed on Perkin-Elmer DSC 8000 calorimeter equipped with Controlled Liquid Nitrogen Accessory LN2 with a heating/cooling rate of 10 K/min. The sample of the mass 43.201 mg was sealed in the aluminum pans. The measurement was performed for the powder sample in the 100 – 315 K temperature range. The helium as purged gas was used.

The emission and excitation spectra were obtained using the FLS1000 Fluorescence Spectrometer from Edinburgh Instruments equipped with 450 W Xenon lamp and R5509-72 photomultiplier tube from Hamamatsu in nitrogen-flow cooled housing for near infrared range detection. To carry out the temperature measurement, the temperature of the sample was controlled using a heating stage from Linkam (0.1 1C temperature stability and 0.1 1C set point resolution) and emission spectra were measured using the 808 nm and 1060 excitation lines from a laser diodes (LD). Luminescence decay profiles were recorded using using using the FLS1000 Fluorescence Spectrometer from Edinburgh Instruments equipped with 450 W Xenon lamp and R5509-72 photomultiplier tube from Hamamatsu in nitrogen-flow cooled housing for near infrared range detection.

*Microscopic spectrally-, spatially- and temporally-resolved measurements*

Microscopic measurements were performed using setup based on Nikon Eclipse Ti-U microscopic body. The excitation laser beam (400 nm), provided by the laser diode (395 nm), entered by the back port of the microscope body and was directed by the dichroic mirror (DAPI-50LP-A, Semrock) to the objective lens (Plan Apo λ 2x, NA=0.10, Nikon) used, both, for sample illumination and collecting its luminescence. The emitted light, collected by the objective lens, after passing through the dichroic mirror, was directed to the side port of the microscopic body and was filtered by the longpass filter (FELH0500, Thorlabs). Next, it was formed by the lens to produce the image of sample luminescence on the entrance slit of the monochromator (Shamrock 500i, Andor) equipped with CCD detector (Newton 920, BEX2-DD, Andor) operating in two modes. In the first one, the entrance slit was fully open (slit of 2500 µm), and the grating assembled inside the monochromator was set in the position of the $0^{th}$ order diffraction mode and directing reflected light to the CCD chip. As a result, an image of luminescence of the illuminated part of the sample was observed (1.5 mm in the vertical direction). In the latter mode the entrance slit was set to 200 μm, and the diffraction grating was operating in the $1^{st}$ diffraction mode. As a result, from each of the points of the sample located along the slits emission spectra were acquired simultaneously forming a 1D hyperspectral and temporal image. In order to improve the signal to noise ratio in this measurements, the binning of the CCD pixels was utilized (1 x 16, where the first parameter describes the horizontal direction of the CCD chip, corresponding to the wavelength of incoming light, while the latter describes the vertical direction, corresponding to the spatial position on the sample). On top of that, these spatio-spectral data were collected in the kinetic mode with signal acquisition time of 1 s per each of the 900 frames, hence providing overall information (luminescence intensity) about spectral (586-594 nm range), spatial (1.5 mm length across the sample) and temporal (900s) evolution of the sample's luminescence.

Next, collected datasets were analysed using home-built Matlab scripts to translate this 4D spectral information into the corresponding temperature distribution over length and over time.